\documentclass[11pt]{iopart}

\usepackage{graphicx}

\usepackage{epstopdf}	

\usepackage[utf8]{inputenc}
\usepackage{hyperref}
\hypersetup{
    colorlinks=false,       
    colorlinks=true,
    linkcolor=blue,          
    citecolor=red,        
    filecolor=magenta,      
    urlcolor=cyan           
}
\usepackage{xcolor}

\newcommand{\frob}[1]{
\left\|#1\right\|_F
}
\newcommand{\ket}[1]{\left|{#1}\right\rangle}

\usepackage{bbold}

\begin{document}
\title[Acc. 2D magn. resoncance spectroscopy using matrix completion]{Accelerated 2D magnetic resonance spectroscopy of single spins using matrix completion}

\author{Jochen Scheuer$^{1}$, Alexander Stark$^{1}$, Matthias Kost$^{2}$, Martin B. Plenio$^{2}$, Boris Naydenov$^{1}$, and Fedor Jelezko$^{1}$}

\address{
$^1$Institut f\"ur Quantenoptik and Center for Integrated Quantum Science and Technology (IQST), Albert-Einstein-Allee 11, Universit\"at Ulm, 89069 Ulm, Germany\\
$^2$Institut f\"ur Theoretische Physik and Center for Integrated Quantum Science and Technology (IQST), Albert-Einstein-Allee 11, Universit\"at Ulm, 89069 Ulm, Germany\\
}

\ead{boris.naydenov@uni-ulm.de}

\renewcommand{\comment}[1]{{\it \color{red} #1 }}

\begin{abstract}
Two dimensional nuclear magnetic resonance (NMR) spectroscopy is one of the major tools for 
analysing the chemical structure of organic molecules and proteins.
Despite its power, this technique requires long measurement times, which, particularly in the
recently emerging diamond based single molecule NMR, limits its application to stable samples.
Here we demonstrate a method which allows to obtain the spectrum by collecting only a small 
fraction of the experimental data. Our method is based on matrix completion which can recover 
the full spectral information from randomly sampled data points.
We confirm experimentally the applicability of this technique by performing two dimensional 
electron spin echo envelope modulation (ESEEM) experiments on a two spin system consisting 
of a single nitrogen vacancy (NV) centre in diamond coupled to a single $^{13}$C nuclear spin.
We show that the main peaks in the spectrum can be obtained with only 10~$\,$\% of the total 
number of the data points.
We believe that our results reported here can find an application in all types of two dimensional 
spectroscopy, as long as the measured matrices have a low rank.
\end{abstract}

\submitto{\NJP as Fast Track Communication}

\maketitle

\section{Introduction}
A key tool in the quest for the determination of the structure of molecules and
proteins is nuclear magnetic resonance spectroscopy (NMR) which has helped to
make fundamental contributions to the advancement of biological sciences.
This is achieved by measuring the magnetic response of molecules in a large
ensemble to sequences of radio frequency pulses. This temporal response is then
mapped to multi-dimensional spectra which encode the dynamical properties of
the system and therefore the interactions between its constituent nuclear spins
\cite{Jeener71,Ernst89}. The information contained in these spectra forms the basis for the
determination of molecular structure. Current NMR schemes are intrinsically
ensemble measurements, both due to the minute size of the nuclear magnetic
moments and the tiny polarization of these nuclear spins at room temperature,
even in very strong magnetic fields. Consequently, NMR can only deliver ensemble
information while the structure and dynamics of individual specimens remain
hidden from observation.

Recent progress in the control of the single electron spin in Nitrogen-vacancy (NV)
centers in diamond offers a new perspective here, as it can make use of optically
detected magnetic resonance \cite{Koehler93,Wrachtrup93} for the detection of material
properties \cite{Gruber97} including minute magnetic fields \cite{Chernobrod05,Degen08,Maze08,Gopi08}. 
Building on this,
recent theoretical investigations \cite{Cai13,Viktor13,Kost2014,Ajoy15} have
suggested that NV centers implanted a few nanometers below the surface should be
able to detect and locate individual nuclear spins above the diamond surface.
Subsequent experimental work has indeed achieved the observation of small clusters
of nuclear spins outside of diamond with a sensitivity that is sufficient to identify
even individual nuclear spins \cite{Mueller14}.

One of the challenges for the determination of the structure of smaller biomolecules
or even entire proteins by means of 2D spectroscopy detected by an NV center is the
considerable amount of data that need to be taken which results in long measurement
times. Indeed, the large amount of required data and the associated long measurement times
represent a challenge that is common to both ensemble NMR and single molecule NMR measurements.

As suggested in \cite{Kost2014} we demonstrate NV sensing experiments on nuclear spins 
using methods from the field of signal processing, particularly matrix completion 
\cite{CandesW08,JFC10}. With this technique we can obtain reliably the spectral
information that is contained in 2D-NMR spectra from a small subset of all accessible
data points (see \cite{HollandBG+11} and \cite{Hoch14} for applications of the related but distinct 
compressive sensing and non-uniform sampling to bulk NMR). The results presented here show that order of magnitude 
reduction in the overall measurement time in NV center based 2D-NMR can be achieved.

In the remainder we briefly introduce matrix completion in Section \ref{MatrixCompletion}.
Then Section \ref{ExperimentalImplementation} presents the application of this method to
concrete experimental data that have been obtained from an NV center interacting with a 
nearby nucleus. The results demonstrate that already a sampling rate of around $10\%$ 
suffices to reconstruct the spectral information reliably. We finish with a brief conclusion 
and outlook concerning the potential of this approach for diamond quantum sensing.

\section{Matrix Completion Method}
\label{MatrixCompletion}

This section serves to introduce briefly the concept of matrix completion, the basic
properties relevant to this work and the specific algorithm that we use for its application
to our experimental data.

A 2D-spectrum encodes the response of a system to a sequence of pulses with varying temporal
separation, denoted by $t_i$ and $\tau_j$, and arrange the result in a data matrix $M(i,j)$.
The 2D-spectrum $S(\omega_1,\omega_2)$ is then obtained as the Fourier transform of both
time coordinates in $M$. In our work we are sampling randomly chosen elements of the matrix $M$
with indices $(i,j)$ drawn from the index set $\Omega$, leading to constraints $X_{ij}=M_{ij}$ 
for $(i,j)\in \Omega$. Matrix completion solves the task of obtaining the missing matrix entries 
of $M$ that have not been measured in experiment. In general this is impossible unless we have 
further knowledge about the matrix $M$, namely that it typically has a low singular value rank 
$r$, i.e. $r\ll n$. Fortunately, this is indeed the case for typical data sets and especially 
those from 2D spectroscopy.

One possible approach to achieve this matrix completion is by solving the minimization problem
\begin{equation}
    \min\,\left[\,tr|X|\, : |X_{ij} -  M_{ij}| < \epsilon\; \mbox{for}\; (i,j)\in \Omega\,\right]
    \label{tracenorm}
\end{equation}
where $tr|X|$ is the trace norm of the matrix $X$ and $\epsilon$ is a given tolerance. Indeed, 
it can be proven that this formulation of the problem achieves the desired aim \cite{Candes2009} 
as the solution of eq. (\ref{tracenorm}) yields the matrix $M$ with high probability if the number 
of sampled elements $|\Omega|={\cal O}(nr\ln n)$ where $r$ is the singular value rank of the 
$n\times n$ matrix $M$ (see \cite{Candes2009,Gross11} for proofs and a rigorous mathematical statement).
This suggests that a computational gain by a factor of order $n/(r\ln n)$ may be achieved
through random sampling in the manner described above (see for example \cite{Kost2014,Almeida12} on
computed 2D-spectroscopy data).

This still leaves us with the task of solving the minimization problem eq.~(\ref{tracenorm}).
In principle, this equation can be rewritten as a semi-definite programme and then
solved employing standard solvers for convex problems. Unfortunately, standard solvers tend
to be limited to relatively small matrix sizes, but fortunately alternatives exist. Indeed, \cite{JFCai2008}
proposed to solve eq.~(\ref{tracenorm}) approximately through the so-called singular value
thresholding (SVT) algorithm \cite{JFCai2008} which permits very large matrices to be treated.
It is this algorithm that we will be using in our work. The SVT-algorithm solves iteratively
the set of equations
\begin{eqnarray}
    Y^{(k-1)} &=& U^{(k-1)}D^{(k-1)} V^{(k-1)}\label{SVT1}\\
    X^{(k)} &=& U^{(k-1)}\max\left(D^{(k-1)}-\tau \mathbb{1}\right)V^{(k-1)}\\
    Y^{(k)} &=& Y^{(k-1)} + \delta_k {\cal P}_{\Omega}(M-X^{(k)})\label{SVT}
\end{eqnarray}
where $({\cal P}_{\Omega}(M))_{ij}=M_{ij}$ for $(i,j)\in \Omega$ and zero otherwise and eq.~(\ref{SVT1}) represents the singular value decomposition of the matrix $Y^{(k-1)}$. $\tau$
and $\delta_k$ are free parameters in the procedure that regulate the soft thresholding
(eq. 4) and the inclusion of the constraints (eq. 5). The choice $\delta_k<2$ ensure provable
convergence and $\tau=5n$ for $n\times n$-matrices represent typical values (see \cite{JFCai2008}).
As a termination criterion of the iteration we employ the condition
\begin{equation}
    \frac{\frob{({\cal P}_{\Omega}(X^{(k)}-M)}}{\frob{({\cal P}_{\Omega}(M)}} < \epsilon
\label{convergence}
\end{equation}
for some $\epsilon$, and $||.||_F$ being the Frobenius norm \cite{JFCai2008}. The algorithm 
employs a singular value decomposition which, for large matrices, can be accelerated considerably
\cite{Halko11,Tamascelli15}. It is also noteworthy that other approaches for solving eq.~(\ref{tracenorm}) 
such as those reported in \cite{Keshavan2009, Dai2009, Balzano2010} may lead to improved
performance and/or stability but for the purposes of this study SVT was sufficient and
recommended itself thanks to its ease of implementation.
In any real-world application, the measured entries of the data matrix will be corrupted
at least by a small amount of noise. Hence the question of the robustness of the matrix
completion approach to fluctuations in the experimental data arises naturally. Reassuringly, 
results have been developed that guarantee that reasonably accurate matrix completion is 
possible from noisy sampled entries \cite{CandesP10}. In that scenario noise can be neglected 
if the relevant spectral information can be still extracted from the low rank approximation 
of $M$, thus implicating a sufficiently large signal-to-noise ratio and leading to the fact 
that noise contribution results in small singular values, which are discarded after applying 
our algorithm.
Hence matrix completion offers three major advantages:
\begin{itemize}
\item Weak noise is directly suppressed by the matrix completion algorithm
\item The spectrum of the system can be recovered from a small subset of all
data e.g. only 10$\,$\% of the total in our examples.
\item In contrast to compressed sensing, the algorithm used here does not
require any additional information, e. g. the sparse basis \cite{Ma13}.
\end{itemize}
The following section will now present the result of the application of the matrix
completion algorithm to concrete experimental data that have been obtained in our
laboratory.

\section{Experimental Implementation}
\label{ExperimentalImplementation}
\subsection{2D ESEEM with a single NV centre}
\begin{figure}[htbp]
\centering
\includegraphics[scale=0.8]{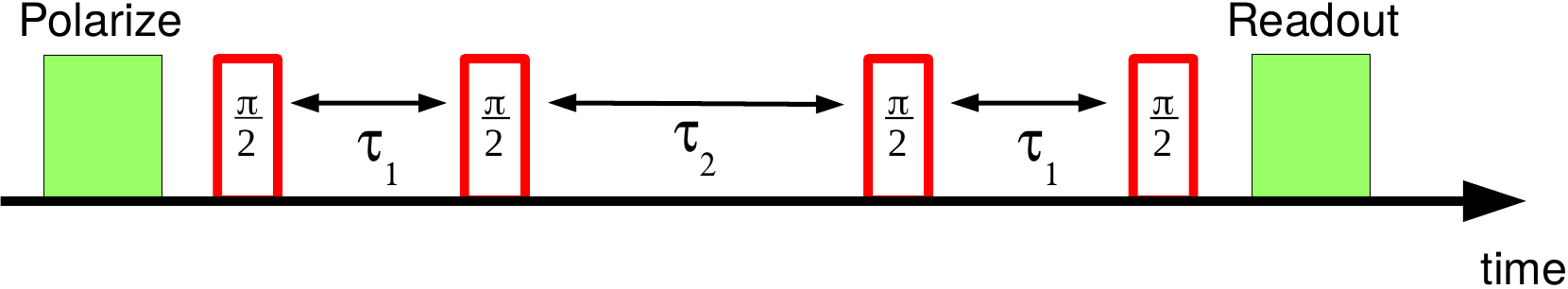}
\caption{Pulse sequence for the two dimensional ESEEM measurement used in our experiments.}
\label{PulseSequence}
\end{figure}

The method of matrix completion has been implemented in 2D optical spectroscopy of Rb 
vapour \cite{Sanders12}.
We use a single NV centre in diamond coupled to a proximal $^{13}$C nuclear spin as a 
test system for the demonstration of the matrix completion protocol.
NVs are optically active point defect centres in the diamond crystal.
Their fluorescence depends on the electron spin number $m_s$ of the triplet ground 
state, allowing to measure the electron spin of single centres.
NVs close to the diamond surface have been used to detect few thousand external protons \cite{Staudacher13, Mamin13} followed later by a demonstration of even single spins sensitivity \cite{Mueller14} leading to nano-scale magnetic resonance imaging \cite{Rugar15, Haeberle15, Walsworth15}. For these type of experiments the data acquisition is quite long due to the low fluorescence emission from single centres.\\
The NV has a triplet ground state (electron spin $S=1$) coupled to the nitrogen nuclear spin ($^{14}$N, $I=1$).
The system can be described by the Hamiltonian:
\begin{equation}
\hat H =  D \hat{S_z^2}+\frac{g\mu_B}{\hbar}\vec{B}\cdot\vec{S}+\vec{S}\cdot\mathbf{A_{\mathrm{^{14}N}}}\cdot\vec{I} \,
\label{NVHamiltonian}
\end{equation}
where $D/2\pi=2.87\,$GHz is the zero field splitting of the ground state,  $g=2.003$ is the Land\'{e} factor, $\mu_B$ is the Bohr magneton, $\vec{B}=B_x\vec{e}_x+B_y\vec {e}_y+B_z\vec{e}_z$ is the applied static magnetic field, $\vec{S}=\hat S_x+\hat S_y+\hat S_z$ and $\vec{I}=\hat I_x+\hat I_y+\hat I_z$ are the electron and nuclear spin operators and $\mathbf{A_{\mathrm{^{14}N}}}$ is the hyperfine interaction tensor.
The $z$ axis is taken to be along the NV crystal axis.
If there is a single $^{13}$C nuclear spin ($I=1/2$) in the proximity, the following term
\begin{equation}
\hat H_{\mathrm{HF^{13}C}} = \vec{S}\cdot\mathbf{A_{\mathrm{^{13}C}}}\cdot\vec{I}
\label{NV13CHamiltonian}
\end{equation}
is added to the spin Hamiltonian \eref{NVHamiltonian}, with $\mathbf{A_{\mathrm{13C}}}$ being the hyperfine interaction tensor to a $^{13}$C nuclear spin.
\noindent One of the simplest 2D NMR experiments consists of three $\pi/2$ pulses and is called correlation spectroscopy (COSY) \cite{Slichter96}.
In our work we use its "equivalent" in the electron spin resonance - the three pulse electron spin echo envelope modulation (ESEEM) pulse sequence (see \cite{Schweiger2001} for more details) shown in \fref{PulseSequence}.

The sequence starts with a laser pulse of about 3 $\mu$s to polarize the NV electron spin in the $\ket{m_s=0}$ state.
Afterwards we apply four $\pi/2$ microwave pulses at times $t=0$, $t=\tau_1$, $t=\tau_1+\tau_2$ and $t=2\tau_1+\tau_2$.
The last pulse is used to transfer the electron spin coherence into population, which is read out by the last laser pulse.
The spin signal is recorded for each pair of ($\tau_1$,$\tau_2$) and then a 2D Fourier transform is performed giving a set of frequencies ($\nu_1$,$\nu_2$).
From this spectrum the number of nuclei coupled to the electron spin and the off diagonal elements of the hyperfine interaction tensor (e.g. proportional to $\hat{S_z}\hat{I_x}$) can be obtained \cite{Schweiger2001}.

\begin{figure}[htbp]
\centering
\begin{tabular}{cc}
\includegraphics[scale=0.59]{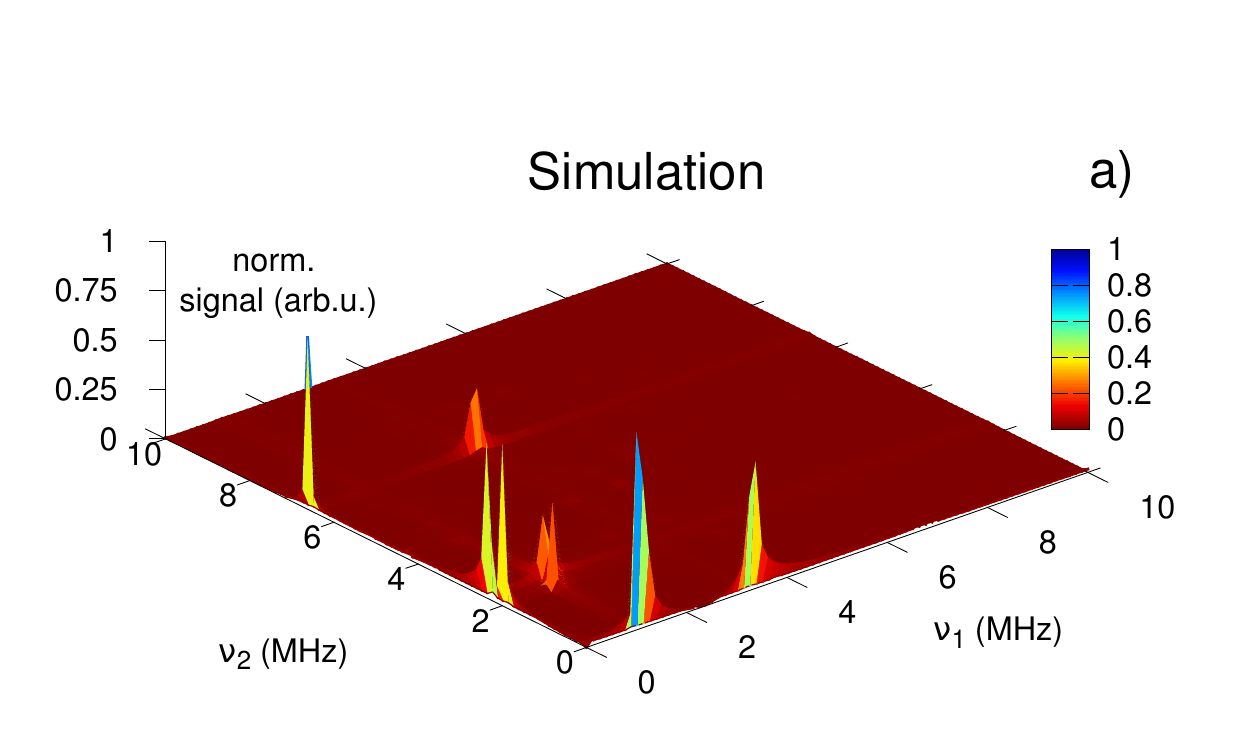} 
\includegraphics[scale=0.59]{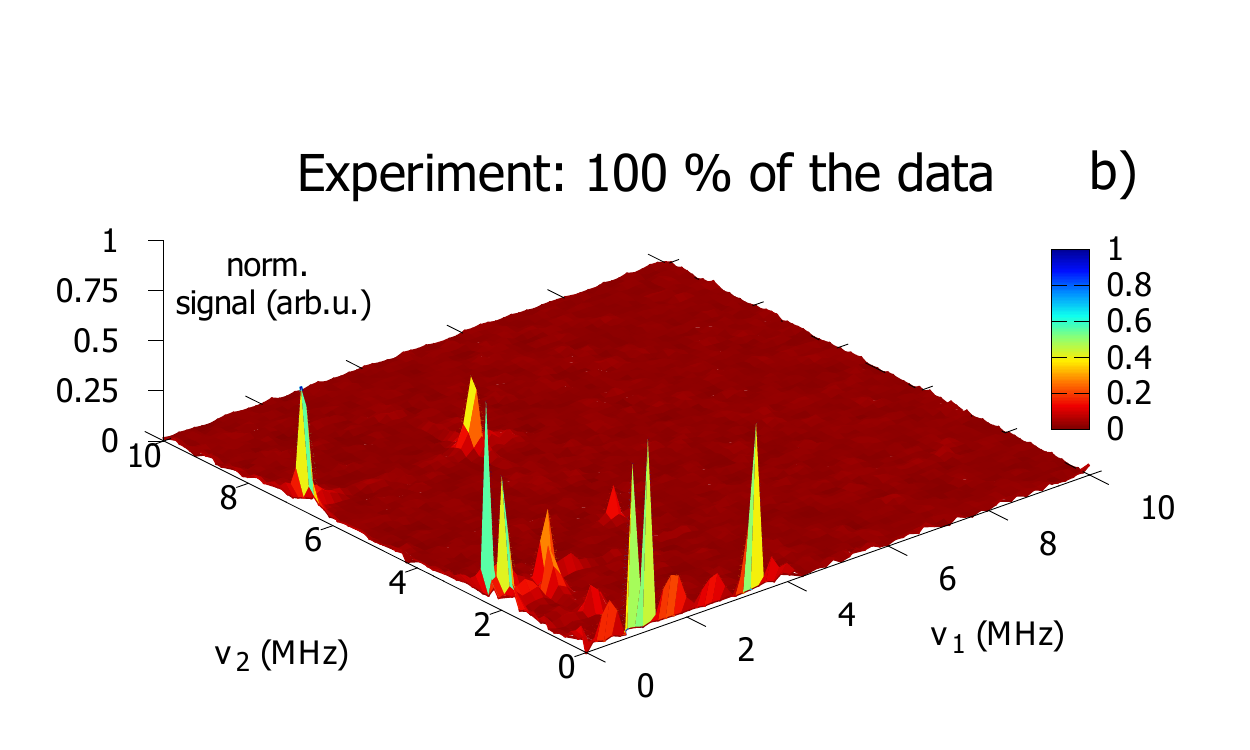}\\
\includegraphics[scale=0.59]{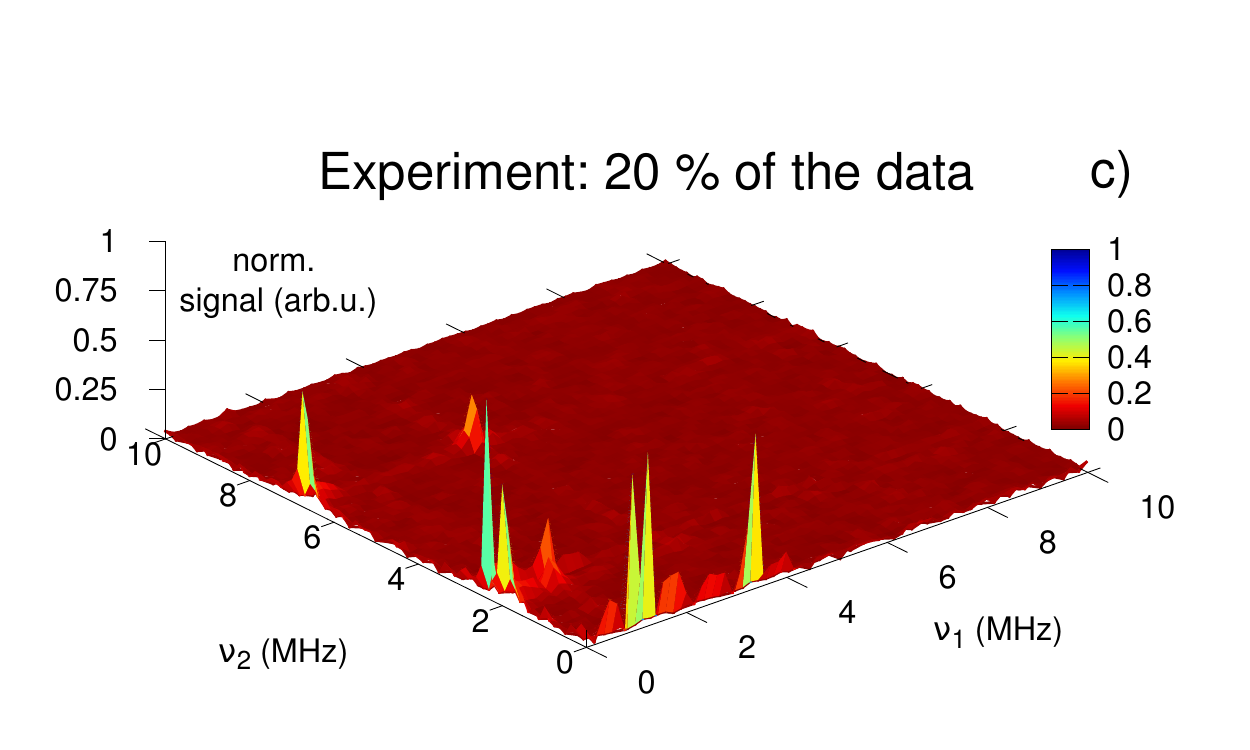}
\includegraphics[scale=0.59]{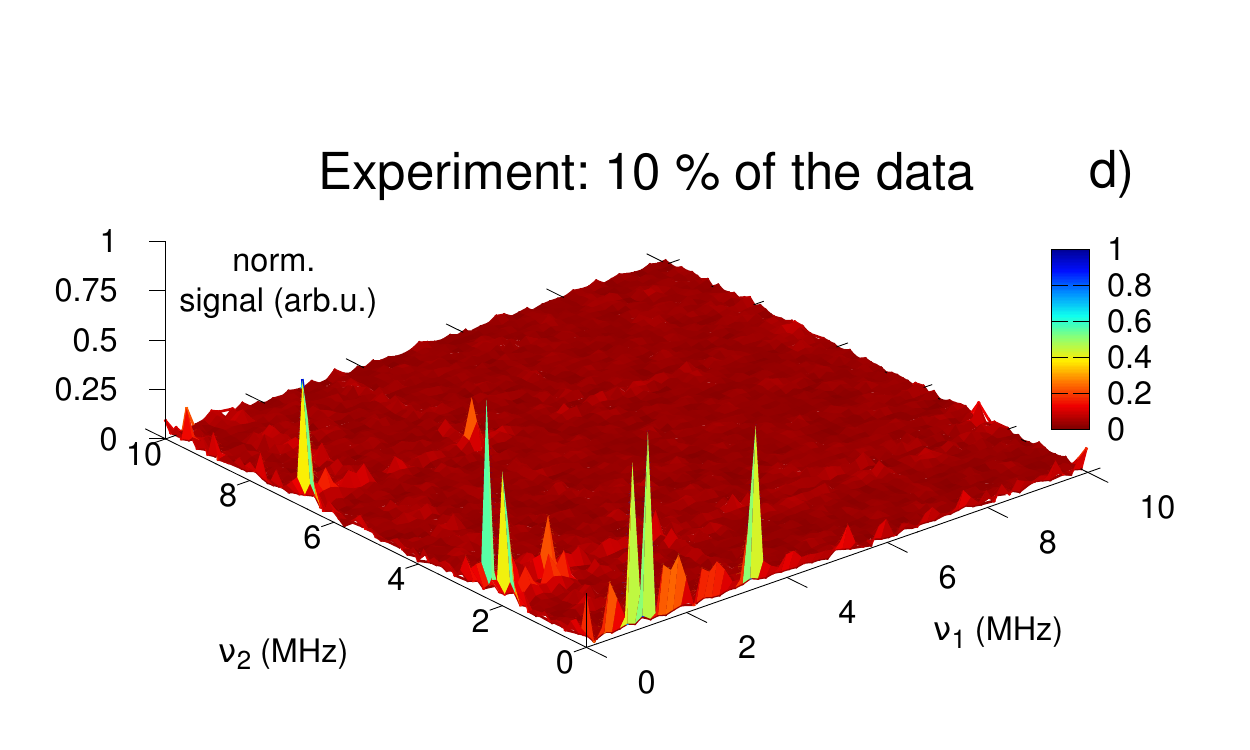}
\end{tabular}
\caption{2D ESEEM simulation and experimental data with a single NV when static magnetic field \mbox{$B_0=100.9\,$G} is applied at an angle of $34.1^{\circ}$.
(a) Simulation using the Hamiltonian~(\ref{NVHamiltonian}).
(b) Fourier transform of the complete set of the experimental data points $S^{\mathrm{full}}_{\mathrm{exp}}$.
Fourier transform of the experimental data after applying matrix completion and using 20$\,$\% $S^{\mathrm{20\,\%}}_{\mathrm{exp}}$ (c) and $S^{\mathrm{10\,\%}}_{\mathrm{exp}}$ 10\,\% (d) of the time domain data. The main peaks are still observed even when 90\,\% of the data is removed!}
\label{Misalignment}
\end{figure}

\noindent We applied this pulse sequence in two different experiments.
In the first measurement we use a single NV without resolvable coupling to $^{13}$C spins.
The system consists of a NV electron and a nitrogen nuclear spins, which are described by the Hamiltonian in \eref{NVHamiltonian}.
If the static magnetic field is aligned with the NV axis, the hyperfine interaction tensor $\mathbf{A_{\mathrm{^{14}N}}}$ is diagonal and there is no ESEEM effect.
In order to introduce artificial "off-diagonal" terms, we apply the static $\vert\vec{B}\vert\approx 100$~G off-axis, at an angle of about $34$ degrees with respect to the $z$ axis.
The expected 2D spectrum $S_{\mathrm{theo}}$ can be simulated by using the Hamiltonian~\eref{NVHamiltonian} and is plotted in \fref{Misalignment}a.
In \fref{Misalignment}b we plot the Fourier transform of the experimental data $S^{\mathrm{full}}_{\mathrm{exp}}$, where we use all collected data points.
The experiment agrees well with the simulation.
In order to demonstrate the performance of the matrix completion method, we use a random mask $\Omega$ to remove a certain part from the full experimental data in the time domain.
After that, we apply matrix completion using the SVT algorithm as described in section~\ref{MatrixCompletion} to recover the full matrix.
A Fourier transform of the matrix obtained with 20\,\% of the initial data points $S^{\mathrm{20\,\%}}_{\mathrm{exp}}$ is shown in \fref{Misalignment}c.
Despite the removal of $80$\,\% of the recorded data, the number of peaks and their positions 
are still present if we compare to \fref{Misalignment}b.
Even if we keep only 10\,\% of the original matrix (see \fref{Misalignment}d), we can still 
recover the relevant spectral information.

\begin{figure}[htbp]
\centering
\begin{tabular}{cc}
\includegraphics[scale=0.59]{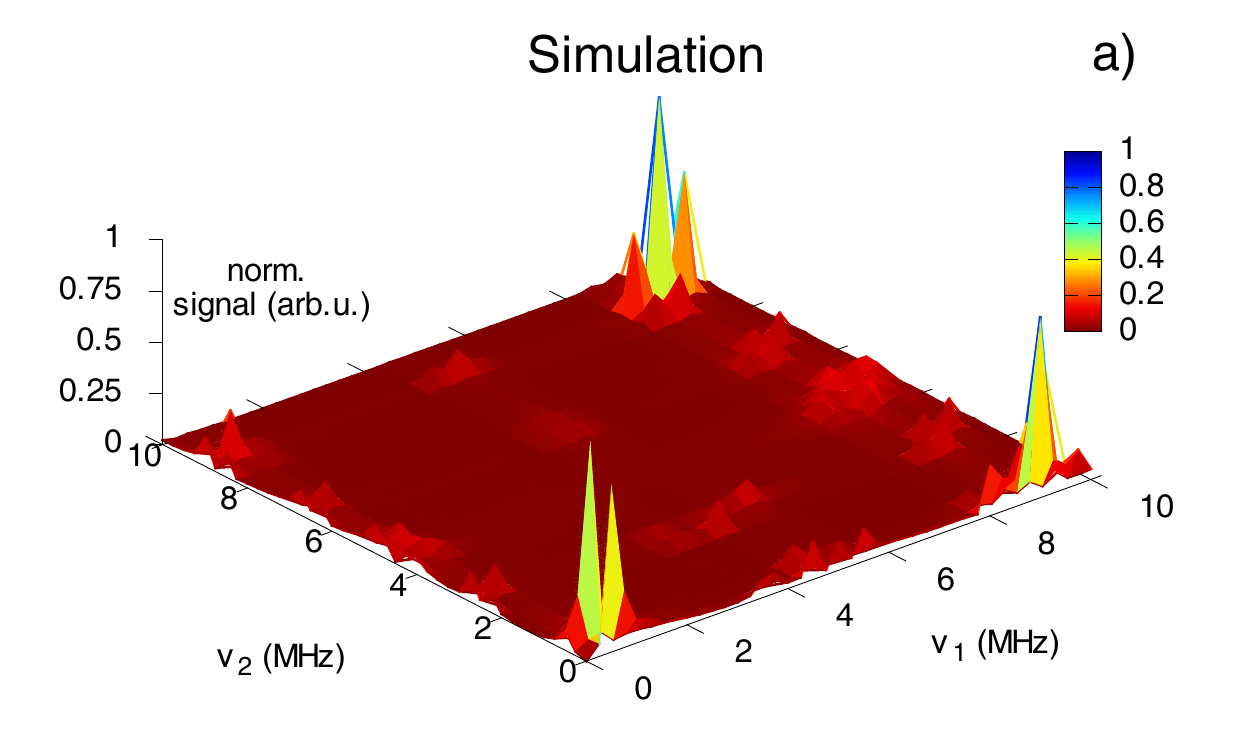} 
\includegraphics[scale=0.59]{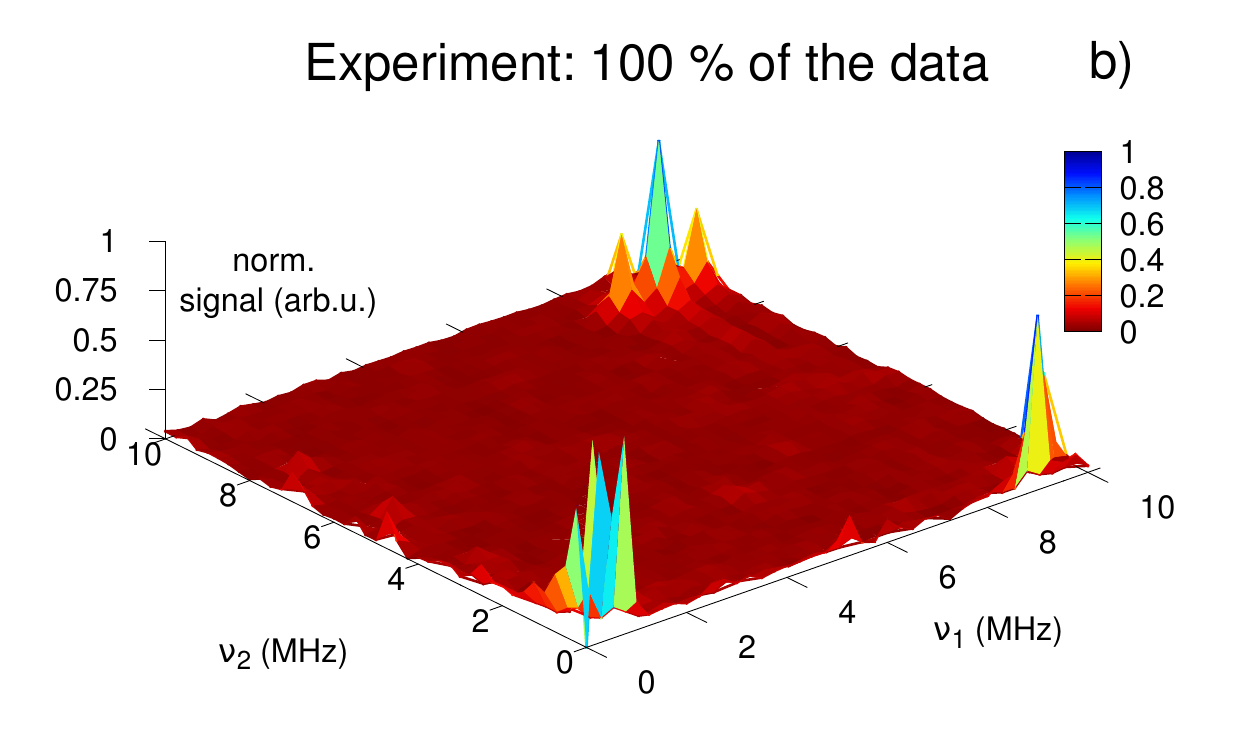}\\ 
\includegraphics[scale=0.59]{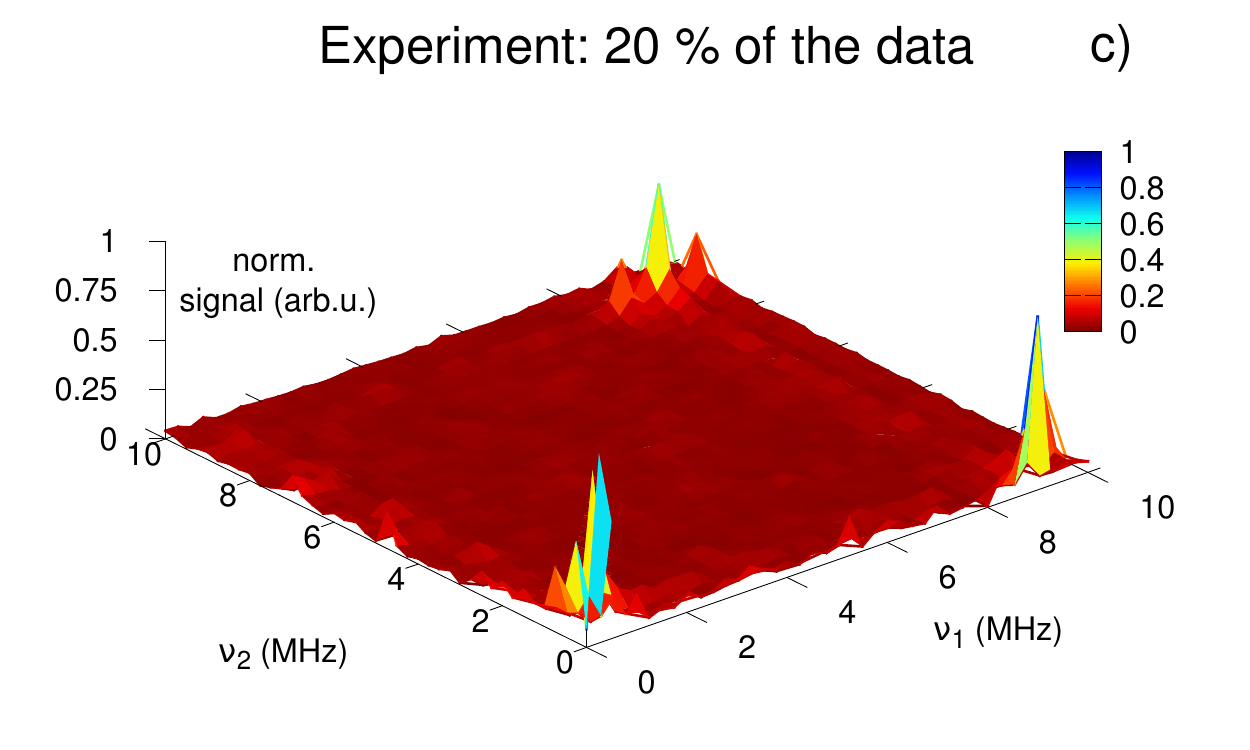} 
\includegraphics[scale=0.59]{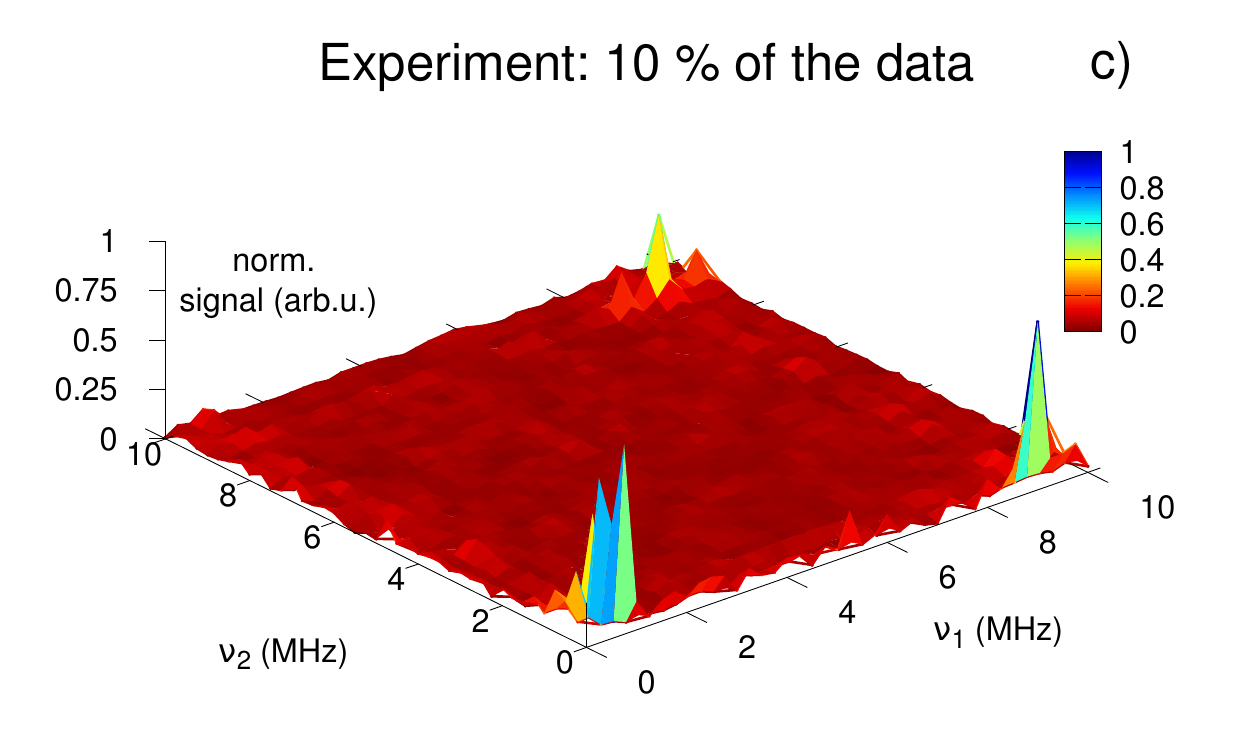} 
\end{tabular}
\caption{2D ESEEM simulation and experimental data with a single NV coupled to a single $^{13}$C nuclear spin.
(a) Simulation using the spin Hamiltonian~(\ref{NVHamiltonian}) and~(\ref{NV13CHamiltonian}).
(b) Fourier transform of the complete set of the experimental data points.
Fourier transform of 20\,\% (c) and 10\,\% (d) of the data in the time domain.
Here we again can recover the spectral information by keeping small amount of the experimental data.}
\label{ESEEM}
\end{figure}

\noindent In the second experiment we localized an NV coupled to a single $^{13}$C spin with a coupling strength of \mbox{$A_{\mathrm{^{13}C}}=9$\,MHz}.
Now, depending on the position of this carbon atom with respect to the NV, there are different hyperfine interaction tensors \cite{Nizovtsev10a, Nizovtsev10b, Nizovtsev14}.
The spectrum can be calculated by using the Hamiltonian~\eref{NVHamiltonian} and \eref{NV13CHamiltonian} by choosing the correct hyperfine interaction tensor.
The simulation is shown in \fref{ESEEM}a.
In these measurements the magnetic field has been aligned along the NV axis.
The 2D Fourier transform of the full data set is shown in \fref{ESEEM}b.
As in the previous experiment, we can still recover the full spectral information 
(cf. \fref{ESEEM}c), if we remove randomly 80\,\% of the data points.
From \fref{ESEEM}d we can conclude that even 10\,\% of the data suffice for the matrix 
completion algorithm to obtain the spectrum. In fact, this factor of ten is what is expected 
from the theory, see section~\ref{MatrixCompletion} and below.

\subsection{Performance of the matrix completion algorithm}
In the following the performance of the matrix completion algorithm will be analysed on our
experimental data.
For this purpose we have to quantify how good we can recover the matrix containing the spectrum, 
when a small number of measurements is performed. We use the experimental data shown in \fref{ESEEM}.
The data is stored in a matrix $M_{\mathrm{red}}$ with reduced number of elements different from zero.
This matrix is then compared with the matrix $M_{\mathrm{tot}}$, obtained by measuring the total 
number of points with all pairs of ($\tau_i$,$\tau_j$).
By using these two matrices, we define the fidelity $F$ of our algorithm as
\begin{equation}
F = 1 - \frac{\frob{M_{\mathrm{tot}}-M_{\mathrm{red}}}^2}{\frob{M_{\mathrm{tot}}}^2} \,.
\label{Eq_Fidelity}
\end{equation}

\begin{figure}[htbp]
\centering
\begin{tabular}{cc}
\includegraphics[scale=1]{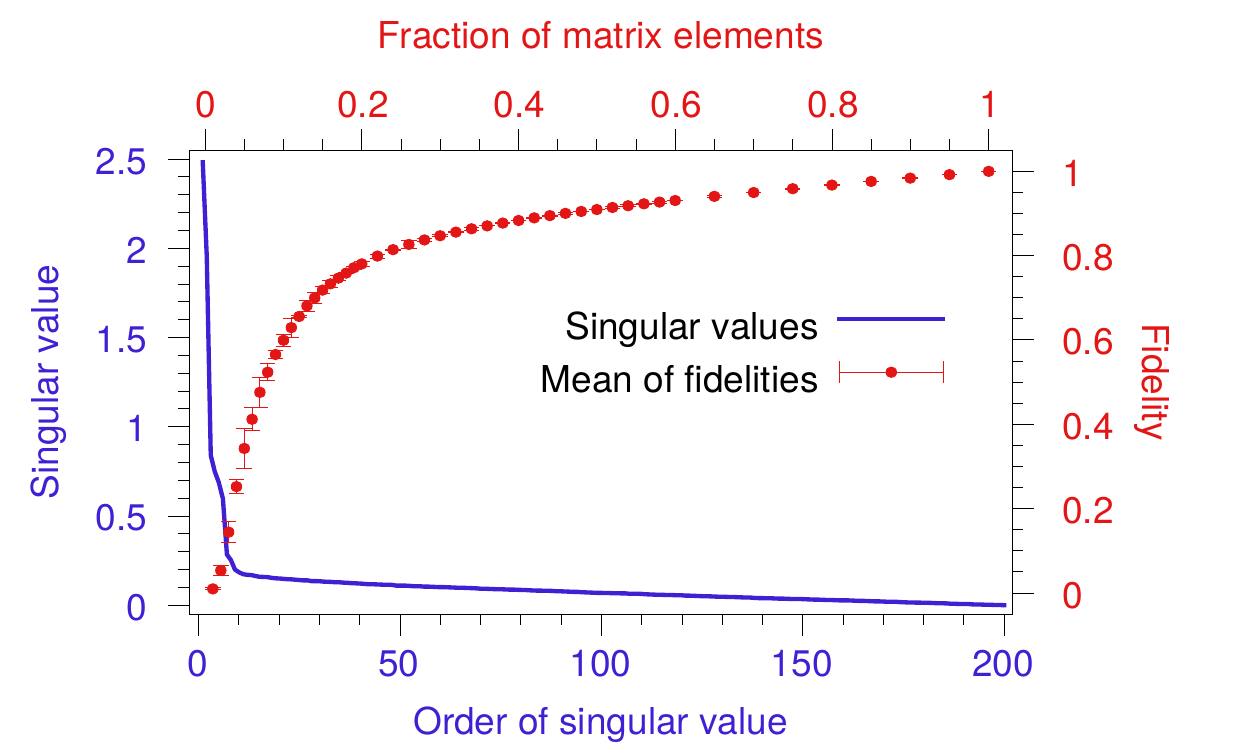}
\end{tabular}
\caption{Mean of fidelities (red dot markers, right axis) of the matrix completion algorithm ($\tau=100$) as a function of the fraction of sampled elements (top axis) taken from the experimental data shown in figure \ref{ESEEM}. The matrix completion algorithm was performed 128 times with each time different random sampling, here the errorbars denote the standard deviation.
The blue curve (left axis) represents the singular values $\sigma_i$ of the measured data (fig. \ref{ESEEM}b) in descending order.}
\label{Fig_Fidelity}
\end{figure}

%
In \fref{Fig_Fidelity} we plot the fidelity as a function of the fraction of the elements of 
the complete matrix (red markers).
In the same plot we show the ordered singular values $\sigma_i$ of the matrix with the full 
number of points $M_{\mathrm{tot}}$.
From the plot we can conclude that most of the spectral information ($>$\,70\,\%) can be 
recovered by using 10\,\% of the elements of $M_{\mathrm{tot}}$, since only these elements 
are significantly larger than zero. This result is consistent with the theoretical limit 
for recovering $M_{\mathrm{tot}}$ given by $nr\ln n$ \cite{Candes2009,Gross11} where we 
can roughly assume $r=4$, which is the number of the peaks in the spectrum.
The spectrum consists of few peaks, while the rest of the matrix elements contain only noise 
and can be discarded.

\begin{figure}[htbp]
\centering
\begin{tabular}{cc}
\includegraphics[scale=1]{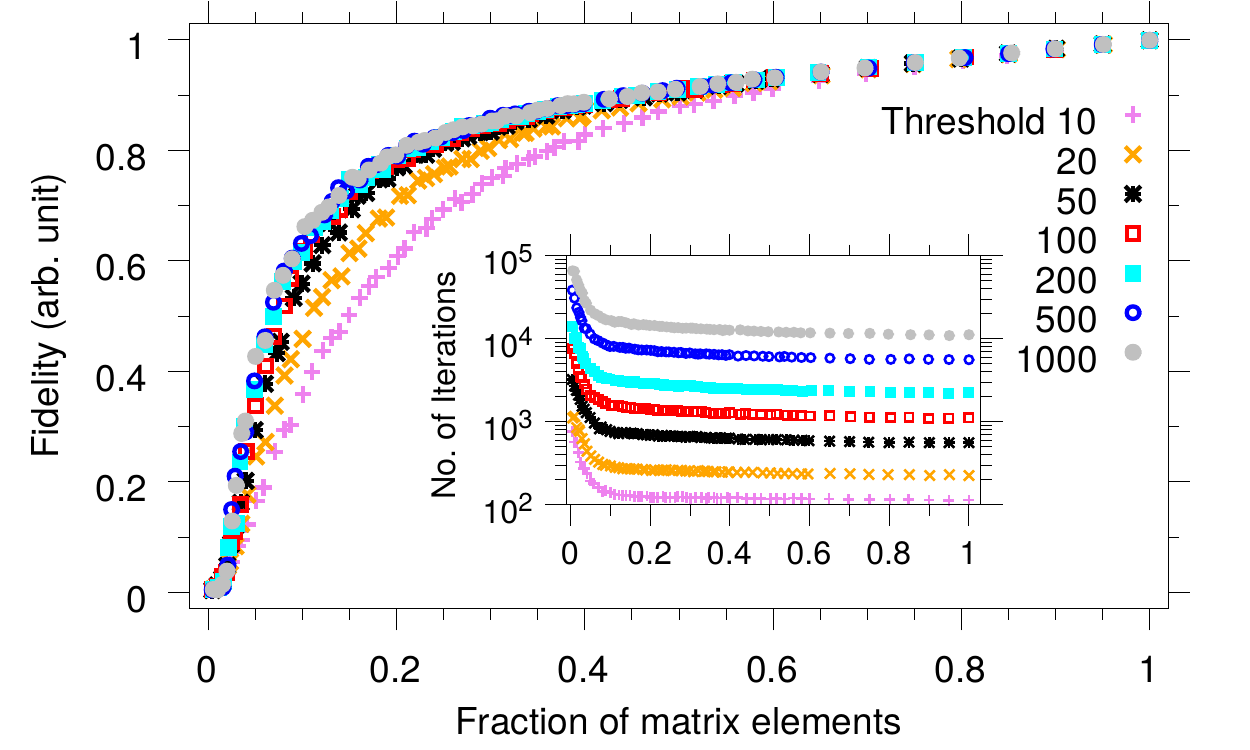}
\end{tabular}
\caption{The fidelity of the matrix completion algorithm as a function of the fraction of the matrix elements at different thresholds $\tau$. Inset: Number of iterations required to run the matrix completion algorithm as a function of the threshold and the fraction of the matrix elements. The performance calculation are based on data shown in figure~\ref{ESEEM}.}
\label{Fig_Fidelity_Threshold}
\end{figure}

\noindent It is interesting to investigate the influence of the threshold parameter $\tau$ on the 
performance of the SVT algorithm and the fidelity of the so determined spectra (see 
\Fref{Fig_Fidelity_Threshold}).
Too small threshold values $\tau$, e.\,g. at $\tau = 10$ or $\tau = 20$ (pink and orange markers), 
lead to low fidelity when less than $60$\,\% of the matrix elements are sampled.
%
%
%
We can achieve higher fidelities $F$ by increasing $\tau$ and we observe saturation around 
$\tau=100$.
That is, for $\tau\gg 100$ we cannot obtain higher significantly higher fidelities, while the 
required computation time (equivalent to the number of iterations) increases which can be seen 
in the inset graph in \fref{Fig_Fidelity_Threshold}. From there we can conclude that with our 
data set thresholds even below the empirically suggested rule $\tau =5\,n \approx 1000$ for our 
case of $n=201$ (see section~\ref{MatrixCompletion} and \cite{JFCai2008}) are sufficient.
A python script can be found in the supplementary information, where the SVT algorithm is implemented, together with the data set from figure~\ref{ESEEM}b.

\section{Conclusions}
In summary, we have demonstrated the application of a method for reconstructing a two 
dimensional ESEEM spectrum, by collecting only small part of the data in the time domain.
With our technique we can obtain the necessary spectral information by measuring 10\,\% of the experimental data points in two different experiments.
By using our method, the measurement time can be shortened by a factor 10 at the same signal-to-noise ratio compared to the conventional experiment.
We believe that the reported results will be useful for any type 2D NMR and ESR spectroscopy and also for magnetic resonance imaging. Our method is particularly useful for single spins experiments, which usually require very long measurement times \cite{Walsworth15, Ajoy15}.

\section*{Acknowledgement}
This work has been supported by DFG (SFB TR21, FOR 1493), Alexander von Humboldt Foundation, Volkswagenstiftung and EU (STREP Project DIADEMS, EQuaM, SIQS, ERC Synergy Grant BioQ).
BN is grateful to the Bundesministerium f\"{u}r Bildung und Forschung (BMBF) for the  ARCHES award.

\vspace{2 cm}
\section*{References}


\begin{thebibliography}{10}

\bibitem{Jeener71}
J.~Jeener.
\newblock {reprinted in {\it NMR and More in Honour of Anatole Abragam, Eds. M.
  Goldman and M. Porneuf}}.
\newblock In {\em Lecture Notes of the Ampere School in Basko Polje, Yugoslavia
  1971}, pages 1--379, 1994.

\bibitem{Ernst89}
R.R. Ernst, G.~Bodenhausen, and A.~Wokaun.
\newblock {\em Principles of Nuclear Magnetic Resonance in One and Two
  Dimensions}.
\newblock Oxford University Press, Oxford, 1989.

\bibitem{Koehler93}
J.~K{\"o}hler, J.A.J.M. Disselhorst, M.C.J.M. Donckers, E.J.J. Groenen,
  J.~Schmidt, and W.E. Moerner.
\newblock {\em Nature}, 363:242, 1993.

\bibitem{Wrachtrup93}
J.~Wrachtrup, C.~von Borczykowski, J.~Bernard, M.~Orrit, and R.~Brown.
\newblock {\em Nature}, 363:244, 1993.

\bibitem{Gruber97}
A.~Gruber, A.~Dr{\"a}benstedt, C.~Tietz, L.~Fleury, J.~Wrachtrup, and C.~von
  Borczyskowski.
\newblock {\em Science}, 276:2012, 1997.

\bibitem{Chernobrod05}
B.M. Chernobrod and G.P. Berman.
\newblock {\em J. Appl. Phys.}, 97:014903, 2005.

\bibitem{Degen08}
C.~L. Degen.
\newblock {\em Appl. Phys. Lett.}, 92:243111, 2008.

\bibitem{Maze08}
J.~R. Maze, P.~L. Stanwix, J.~S. Hodges, S.~Hong, J.~M. Taylor, P.~Cappellaro,
  L.~Jiang, M.~V.~Gurudev Dutt, E.~Togan, A.~S. Zibrov, A.~Yacoby, R.~L.
  Walsworth, and M.~D. Lukin.
\newblock {\em Nature}, 455:644.

\bibitem{Gopi08}
G.~Balasubramanian, I.~Y. Chan, R.~Kolesov, M.~Al-Hmoud, J.~Tisler, C.~Shin,
  C.~Kim, A.~Wojcik, P.~R. Hemmer, A.~Krueger, T.~Hanke, A.~Leitenstorfer,
  R.~Bratschitsch, F.~Jelezko, and J.~Wrachtrup.
\newblock {\em Nature}, 455:648, 2008.

\bibitem{Cai13}
J.-M. Cai, F.~Jelezko, M.~B. Plenio, and A.~Retzker.
\newblock {\em New J. Phys.}, 15:013020, 2013.

\bibitem{Viktor13}
V.~S. Perunicic, L.~T. Hall, D.~A. Simpson, C.~D. Hill, and L.~C.~L.
  Hollenberg.
\newblock {\em Phys. Rev. B}, 89:054432, 2014.

\bibitem{Kost2014}
M.~Kost, J.-M. Cai, and M.B. Plenio.
\newblock {\em http://arxiv.org/abs/1407.6262}, 2014.

\bibitem{Ajoy15}
A.~Ajoy, U.~Bissbort, M.D. Lukin, R.~L. Walsworth, and P.~Cappellaro.
\newblock {\em Phys. Rev. X}, 5:011001, 2015.

\bibitem{Mueller14}
C.~M{\"u}ller, X.~Kong, J.-M. Cai, K.~Melentijevic, A.~Stacey, M.~Markham,
  D.~Twitchen, J.~Isoya, S.~Pezzagna, J.~Meijer, J.F. Du, M.B. Plenio,
  B.~Naydenov, L.P. McGuinness, and F.~Jelezko.
\newblock {\em Nat. Commun.}, 5:4703, 2014.

\bibitem{CandesW08}
E.J. Candes and M.~B. Wakin.
\newblock {\em IEEE Signal Process. Mag.}, 25:21, 2008.

\bibitem{JFC10}
J.-F. Cai, E.~J. Candes, and Z.~Shen.
\newblock {\em SIAM J. on Optimization}, 20:1956, 2010.

\bibitem{HollandBG+11}
D.J. Holland, M.J. Bostock, L.F. Gladden, and D.~Nietlispach.
\newblock {\em Angew. Chem. Int. Ed.}, 50:6548, 2011.

\bibitem{Hoch14}
M.~Mobli and J.~C. Hoch.
\newblock {\em Prog. Nucl. Magn. Reson. Spectrosc.}, 83:21, 2014.

\bibitem{Candes2009}
E.~J. Cand\`{e}s and B.~Recht.
\newblock {\em Foundations of Computational Mathematics}, 9:717, 2009.

\bibitem{Gross11}
D.~Gross.
\newblock {\em IEEE Trans. Inf. Theory}, 57:1548, 2011.

\bibitem{Almeida12}
J.~Almeida, J.~Prior, and M.P. Plenio.
\newblock {\em Journal of Physical Chemistry Letters}, 3:2692, 2012.

\bibitem{JFCai2008}
J.-F. Cai, E.~J. Candès, and Z.~Shen.
\newblock {\em SIAM J. on Optimization}, 20:1956, 2008.

\bibitem{Halko11}
N.~Halko, P.~Martinsson, and J.~Tropp.
\newblock {\em SIAM Review}, 53:217, 2011.

\bibitem{Tamascelli15}
D.~Tamascelli, R.~Rosenbach, and M.B. Plenio.
\newblock {\em http://arxiv.org/abs/1504.00992}, 2015.

\bibitem{Keshavan2009}
R.H. Keshavan, A.~Montanari, and S.~Oh.
\newblock {\em IEEE Trans. Inf. Theory}, 56:2980, 2010.

\bibitem{Dai2009}
W.~Dai and O.~Milenkovic.
\newblock {\em IEEE Trans. Signal Process.}, 59:3120, 2011.

\bibitem{Balzano2010}
L.~Balzano, R.~Nowak, and B.~Recht.
\newblock {\em http://arxiv.org/abs/1006.4046}, 2010.

\bibitem{CandesP10}
E.~J. Candes and Y.~Plan.
\newblock {\em Proc. IEEE}, 98:925, 2010.

\bibitem{Ma13}
D.~Ma, V.~Gulani, N.~Seiberlich, K.~Liu, J.~L. Sunshine, J.~L. Duerk, and M.~A.
  Griswold.
\newblock {\em Nature}, 495:187, 2013.

\bibitem{Sanders12}
J.~N. Sanders, S.~K. Saikin, S.~Mostame, X.~Andrade, J.~R. Widom, A.~H. Marcus,
  and A.~Aspuru-Guzik.
\newblock {\em J. Phys. Chem. Lett.}, 3:2697, 2012.

\bibitem{Staudacher13}
T.~Staudacher, F.~Shi, S.~Pezzagna, J.~Meijer, J.~Du, C.~A. Meriles,
  F.~Reinhard, and J.~Wrachtrup.
\newblock {\em Science}, 339:561, 2013.

\bibitem{Mamin13}
H.~J. Mamin, M.~Kim, M.~H. Sherwood, C.~T. Rettner, K.~Ohno, D.~D. Awschalom,
  and D.~Rugar.
\newblock {\em Science}, 339:557, 2013.

\bibitem{Rugar15}
D.~Rugar, H.~J. Mamin, M.~H. Sherwood, M.~Kim, C.~T. Rettner, K.~Ohno, and
  D.~D. Awschalom.
\newblock {\em Nature Nanotech.}, 10:120, 2015.

\bibitem{Haeberle15}
T.~H\"{a}berle, D.~Schmid-Lorch, F.~Reinhard, and J.~Wrachtrup.
\newblock {\em Nature Nanotech.}, 10:125, 2015.

\bibitem{Walsworth15}
S.~J. DeVience, L.~M. Pham, I.~Lovchinsky, A.~O. Sushkov, N.~Bar-Gill,
  C.~Belthangady, F.~Casola, M.~Corbett, H.~Zhang, M.~Lukin, H.~Park,
  A.~Yacoby, and R.~L. Walsworth.
\newblock {\em Nature Nanotech.}, 10:129, 2015.

\bibitem{Slichter96}
C.~Slichter.
\newblock {\em Principles of Magnetic Resonance}.
\newblock Springer-Verlag, 1996.

\bibitem{Schweiger2001}
A.~Schweiger and G.~Jeschke.
\newblock {\em Principles of pulse electron paramagnetic resonance}.
\newblock Oxford University Press, 2001.

\bibitem{Nizovtsev10a}
A.~P. Nizovtsev, S.~Ya. Kilin, V.~A. Pushkarchuk, A.~L. Pushkarchuk, and S.~A.
  Kuten.
\newblock {\em Optics and Spectroscopy}, 108:230, 2010.

\bibitem{Nizovtsev10b}
A.~P. Nizovtsev, S.~Ya. Kilin, P.~Neumann, F.~Jelezko, and J.~Wrachtrup.
\newblock {\em Optics and Spectroscopy}, 108:239, 2010.

\bibitem{Nizovtsev14}
A.~P. Nizovtsev, S.~Ya. Kilin, A.~L. Pushkarchuk, V.~A. Pushkarchuk, and
  F~Jelezko.
\newblock {\em New J. Phys.}, 16:083014, 2014.

\end{thebibliography}

\end{document}